\documentclass[journal]{IEEEtran}

\usepackage[cmex10]{amsmath}
\usepackage{amsfonts, amssymb}

\newtheorem{theorem}{Theorem}[section]
\newtheorem{lemma}[theorem]{Lemma}
\newtheorem{proposition}[theorem]{Proposition}
\newtheorem{corollary}[theorem]{Corollary}

\newtheorem{definition}[theorem]{Definition}
\newtheorem{example}[theorem]{Example}
\newtheorem{condition}{Condition}

\newcommand{\bern}{{\rm{Bern}}}
\newcommand{\bino}{{\rm{Bin}}}

\newcommand{\ep}{{\mathbb {E}}}
\newcommand{\pr}{{\mathbb {P}}}
\newcommand{\qr}{{\mathbb {Q}}}
\newcommand{\vc}[1]{{\mathbf #1}}
\newcommand{\vd}[1]{{\boldsymbol #1}}

\newcommand{\EE}{{\mathcal{E}}}
\newcommand{\JJ}{{\mathcal{J}}}
\newcommand{\blah}[1]{}

\begin{document} 

\title{Monotonicity, thinning and discrete versions of the Entropy Power Inequality}
\author{Oliver~Johnson and~Yaming~Yu~\IEEEmembership{Member,~IEEE}
\thanks{Oliver Johnson is with the 
Statistics Group, Department of Mathematics, University of Bristol, University Walk, Bristol, BS8 1TW, UK. e-mail {\tt o.johnson@bris.ac.uk}.
Yaming Yu is with the 
Department of Statistics, University of California, Irvine, CA 92697, USA. e-mail {\tt yamingy@uci.edu}. 
The collaboration between Oliver Johnson and Yaming Yu is supported by an EPSRC
Grant, reference {\tt EP/H002200/1}.}}
\date{\today}
\maketitle

\begin{abstract} \noindent
 We consider the entropy of
sums of independent discrete random variables, in analogy with
Shannon's Entropy Power Inequality, where equality holds for normals.
In our case, infinite divisibility  suggests
that equality should hold for Poisson variables. We show that 
some natural analogues of the Entropy Power Inequality do not in fact hold,
but propose an alternative formulation which does always hold. 
The key to many proofs of Shannon's Entropy Power Inequality is
the behaviour of entropy on scaling of continuous random variables.
We believe that R\'{e}nyi's operation of thinning discrete
random variables plays a similar role to scaling, and
give a sharp bound on how the entropy of
ultra log-concave random variables behaves on thinning. 
In the spirit of the monotonicity results established by Artstein, Ball,
Barthe and Naor, we prove a stronger version of concavity of entropy,
which implies a strengthened form of our discrete Entropy Power Inequality.
\end{abstract}



\section{Review of previous work}

It is natural to consider
the entropy of the sum of independent random variables, for example
in proving theoretical results concerning the Central Limit Theorem or in 
practical models of information transmission involving addition of noise to the signal.

Pedagogically speaking,
the entropy $H$ of discrete random variables usually comes first, with the 
differential entropy $h$ of continuous
random variables coming later.  However, results from functional analysis imply
properties of the differential entropy  which do not yet have discrete 
counterparts. For example
Shannon \cite{shannon} stated Theorem \ref{thm:epi}, known as the Entropy
Power Inequality (EPI),
which was later rigorously proved 
by Stam \cite{stam} and by Blachman \cite{blachman} using an argument 
based on the heat equation. Write $E(t) = \frac{1}{2} \log(2 \pi e t)$ for 
the entropy of a  Gaussian random
variable with finite variance $t$, and define $v(X) = E^{-1}(h(X)) = 
e^{2h(X)}/(2 \pi e)$ 
for the entropy power of random variable $X$ with differential entropy $h(X)$.
(We use $\log$ to represent the natural logarithm throughout this paper).
\begin{theorem}[EPI] \label{thm:epi}
For independent continuous $X$ and $Y$, the
sum $X+Y$ satisfies
\begin{equation} \label{eq:epi} v(X+Y) \geq v(X) + v(Y),\end{equation}
with the only non-trivial case of equality being when $X$ and $Y$ are Gaussian.
\end{theorem} 

 A key role is played in many proofs of Theorem \ref{thm:epi}
by the operation of scaling of continuous random variables, using the fact that
  for
any $\alpha$, 
\begin{equation} \label{eq:vscaling} v(\sqrt{\alpha} X) = \alpha v(X). \end{equation}
One major 
contribution of this paper is Theorem \ref{thm:rtepi}
below, which shows that a one-sided version of (\ref{eq:vscaling}) holds
for discrete random variables. In this case, the operation of scaling
is replaced by the thinning operation introduced by R\'{e}nyi \cite{renyi4}.

As is implicit in the work of Verd\'{u} and Guo \cite{verdu},  
Theorem \ref{thm:epi} can be rephrased in terms of scalings, in the form of
Corollary \ref{cor:epilike} below.
Lieb \cite{lieb2} and
Dembo, Cover and Thomas \cite{dembo} prove the Entropy Power Inequality by working 
with the R\'{e}nyi entropy (a generalization of Shannon's quantity).
They use
properties of $p$-norms on convolution given by Beckner's sharp form \cite{beckner} of the Young
inequality. Using a particular parameterization, they show 
that this Young inequality implies that the differential entropy 
is concave with respect to normalized linear combinations, that is,
for any $0 \leq \alpha \leq 1$:
\begin{equation} \label{eq:entcon}
h(\sqrt{\alpha} X + \sqrt{1-\alpha} Y) \geq \alpha h(X) + (1-\alpha) 
h(Y).
\end{equation}
The papers \cite{dembo,lieb2} show that
(\ref{eq:entcon}) is equivalent to the
Entropy Power Inequality. The form
of $\alpha$ used in this proof suggests 
the following result: 
\begin{corollary} \label{cor:epilike}
 Given independent random variables $X$ and $Y$ with finite and
non-zero entropy power,
there exist $X^*$ and $Y^*$ such that $X = \sqrt{\alpha} X^*$ and 
$Y = \sqrt{1-\alpha}
Y^*$ for some $0 < \alpha < 1$, and such that $h(X^*) = h(Y^*)$. 
The Entropy Power 
Inequality Theorem \ref{thm:epi} is equivalent to the fact that 
\begin{equation} \label{eq:epirephrase} h(X+Y) \geq h(X^*),\end{equation}
with equality if and only if $X$ and $Y$ are Gaussian.
\end{corollary}
\begin{IEEEproof}
Applying (\ref{eq:vscaling}) and taking $\alpha = v(X)/(v(X) + v(Y))$
ensures that $X^* = X/\sqrt{\alpha}$ and $Y^* = Y/\sqrt{1-\alpha}$ have the
property that $v(X^*) = v(Y^*) = v(X) + v(Y)$.

{\bf Assume (\ref{eq:epirephrase})}.
Since $h(X+Y) \geq h(X^*)$, applying $E^{-1}$ to both sides we deduce that
$v(X+Y) \geq v(X^*)$, which equals $v(X) + v(Y)$, so that the EPI
(\ref{eq:epi}) holds.

{\bf Assume (\ref{eq:epi})}.
Since $v(X+Y) \geq v(X) + v(Y) = v(X^*)$, so applying $E$ to both sides,
we deduce (\ref{eq:epirephrase}).
\end{IEEEproof}

It is natural to conjecture that there should be a version of the EPI for 
discrete entropies $H$.
We will show in Theorem \ref{thm:epilike} that an equivalent of this
rephrased EPI does hold for discrete variables, whereas in Section
\ref{sec:failures} we show that some other apparently natural versions
of Theorem \ref{thm:epi} in fact fail.

In the context of sums of independent continuous random variables,
Artstein, Ball, Barthe and Naor \cite{artstein}
proved a stronger type of result, referred to as monotonicity. Alternative proofs were later given by
Tulino and Verd\'{u} \cite{tulino}
and by Madiman and Barron \cite{madiman}. For example, 
Theorem 2 of 
\cite{artstein} gives the following:
\begin{theorem} \label{thm:hmonotone}
Given independent continuous random variables $X_i$ with
finite variance, for any positive $\alpha_i$ such that $\sum_{i=1}^{n+1} \alpha_i =1$,
 writing $\alpha^{(j)} = \sum_{i \neq j} \alpha_i = 1-\alpha_j$, then
$$ n h \left( \sum_{i=1}^{n+1} \sqrt{\alpha_i} X_i \right) \geq \sum_{j=1}^{n+1}
\alpha^{(j)} h\left( \sum_{i \neq j} \sqrt{\alpha_i/\alpha^{(j)}} X_i \right)
.$$
\end{theorem}

This is called monotonicity since, choosing $\alpha_i = 1/(n+1)$,
it implies that for 
independent and identically distributed $X_i$, the entropy of the normalized
sum $h \left( \sum_{i=1}^{n} X_i/\sqrt{n} \right)$ is monotone increasing
in $n$. Equivalently writing $d(X) = D(X \| \phi_{\lambda_X, \sigma^2_X})$
for the relative entropy from $X$ to a normal of the same mean and variance,
the relative entropy of the normalized
sum $d \left( \sum_{i=1}^{n} X_i/\sqrt{n} \right)$ is  monotone decreasing
in $n$. 

The other major contribution of this paper is Theorem \ref{thm:hmon},
which establishes a discrete analogue of Theorem \ref{thm:hmonotone}.
Such monotonicity results as Theorem \ref{thm:hmonotone}
imply  strengthened Entropy Power
Inequalities.
By choosing 
\begin{equation} \label{eq:alphachoice}
 \alpha^{(l)} = \frac{n v \left( \sum_{i \neq l} Y_i \right)}{
\sum_{j=1}^{n+1}
 v \left( \sum_{i \neq j} Y_i \right)},\end{equation}
(in the case that all $\alpha^{(l)} \leq 1$; if not, the result is automatic)
Artstein et al. \cite{artstein} showed that their Theorem \ref{thm:hmonotone} 
implies the following extension of the EPI,
Theorem \ref{thm:epi}:
\begin{theorem} \label{thm:monotone}
Given independent continuous random variables $Y_i$ with
finite variance, the entropy powers satisfy
$$ n v \left( \sum_{i=1}^{n+1} Y_i \right) \geq \sum_{j=1}^{n+1}
 v \left( \sum_{i \neq j} Y_i \right).$$
\end{theorem}

We observe that this strengthened EPI, Theorem \ref{thm:monotone}, can be
expressed in a similar way to Corollary \ref{cor:epilike}. That is, given
independent random variables $Y_i$, if there exist $\alpha_i$ such that
$\sum_{i=1}^{n+1} \alpha_i =1$ and 
$Y_i^* = Y_i/\sqrt{\alpha_i}$ have entropies such that
$h \left( (\sum_{i \neq j} \sqrt{\alpha_i} Y_i^*)/\sqrt{\alpha^{(j)}} \right) = h^*$ 
are constant in $j$, then 
\begin{equation} \label{eq:rephraseepi}
h \left( \sum_{i=1}^{n+1} Y_i \right) \geq h^*.
\end{equation}
This again follows by observing that for each $j$,  (\ref{eq:vscaling})
implies that
$v( \sum_{i \neq j} Y_i ) = v^* \alpha^{(j)} = e^{2h^*} \alpha^{(j)}/(2 \pi e)$, 
so summing over $j$, the RHS of Theorem \ref{thm:monotone} is
equal to $e^{2h^*} n/(2 \pi e)$, and the result follows. Note that in this
case, the choice of $\alpha^{(l)}$ again coincides with that given by 
(\ref{eq:alphachoice}). In Theorem \ref{thm:discepilike}, we prove a
discrete version of this result.

The structure of the remainder of the paper is as follows.  In Section
\ref{sec:thinning} we introduce the thinning operation, and describe the
resulting analogues of the EPI, Theorem \ref{thm:epi}.
In Section \ref{sec:monotonicity} we show how these results can extend
to provide monotonicity results.
In Section
\ref{sec:failures} we 
discuss two natural versions of the EPI which are
not true.
In the self-contained Appendices \ref{sec:isop}
and \ref{sec:proofmon},
we prove the two main results of the paper, namely the scaling result
Theorem \ref{thm:rtepi} and the monotonicity result Theorem
\ref{thm:hmon}. Although these results are related, they are proved 
independently, the first using a semigroup argument similar to
that in \cite{johnson21} and the second using
an examination of certain Hessian terms, and previous results from
\cite{yu2}. 

There has been considerable interest in proving an Entropy Power Inequality
for discrete random variables.
Some authors \cite{shamai, witsenhausen2,
wyner8, wyner9} have focused on replacing the operation
of integer addition $+$ by modulo 2 addition
 $\oplus$, and obtained similar results in that case. 
As in \cite{kontoyiannis3}, we prefer to retain $+$ as integer addition.
Harremo\"{e}s and Vignat \cite{harremoes3} proved that (\ref{eq:epi})
holds when $X$ and $Y$ are any independent
binomial $\bino(n_X,1/2)$ and $\bino(n_Y,1/2)$ random variables,
on redefining $v(X) = e^{2H(X)}/(2 \pi e)$ (simply 
replacing differential entropies $h$
by discrete entropies $H$).
We prefer to conjecture that the discrete version of the Entropy Power Inequality
should be expressed differently, using the entropy of the Poisson
distribution.

\section{Entropy and thinning} \label{sec:thinning}

Recent work of
Harremo\"{e}s, Johnson and Kontoyiannis \cite{johnsonc2, johnson26} shows that, in
many senses related to Information Theory, the equivalent of scaling continuous random variables
by a factor of $\sqrt{\alpha}$ is the thinning operation $T_{\alpha}$ on discrete random
variables, as
introduced by R\'{e}nyi \cite{renyi4}.  
\begin{definition} \label{def:thin} 
The  $\alpha$-thinned version of random variable $Y$ is given by the random sum
$T_{\alpha} Y = \sum_{i=1}^{Y} B_{i}$, 
where the $B_{1},B_{2}\ldots$ are IID Bernoulli $\bern(\alpha)$, 
all independent of $Y$.
\end{definition}

We write $\EE(t) = H( \Pi_t )$, an
increasing concave function, for the entropy of a Poisson
random variable $\Pi_t$ of mean $t$, and define an analogue of the 
entropy power as $V(X) = \EE^{-1}(H(X))$.
Theorem 2.5 of \cite{johnson21} proves 
that $\Pi_{\lambda_X}$
maximises entropy within the class of ultra log-concave (ULC) random variables $X$
(see below) of given mean $\lambda_X$, or that $V(X) \leq \lambda_X$.
We investigate the entropy of sums in the context of 
this restricted ULC class. 
\begin{definition} \label{def:ulc}
The ULC random variables are those whose probability mass functions $P$ satisfy
$$ i P(i)^2 \geq (i+1) P(i+1) P(i-1), \mbox{ \; \; for all $i \geq 1$}.$$
\end{definition}

  The ULC class includes the Poisson family and Bernoulli sums.  
This class was introduced in combinatorics \cite{liggett}, \cite{pemantle}, 
a context in which the Bernoulli random variables are
a natural fundamental building block.

The results outlined in \cite{johnsonc2,johnson26} suggest an
equivalence between scaling
by $\sqrt{\alpha}$ and thinning by $\alpha$. This idea has developed 
with the fact that for discrete random variables, 
a natural equivalent of  (\ref{eq:entcon}) is given by the
following
Thinned Entropy Concavity Inequality, 
proved by Yu and Johnson in \cite{johnsonc6}, extending results in 
\cite{yu2}, and now a consequence of the more general Theorem \ref{thm:hmon}
below.
\begin{theorem}[TECI] \label{thm:teci}
For independent ULC random variables
$X$ and $Y$, for any $0 \leq \alpha \leq 1$
\begin{equation} \label{eq:entcon2}
H(T_{\alpha} X+ T_{1-\alpha} Y) \geq \alpha H(X) 
+ (1-\alpha) H(Y).
\end{equation}
\end{theorem}
 
For $0 < \alpha < 1$, examination of the proof shows that equality
holds if and only if $X$ and $Y$ are Poisson with the same mean. 

One major contribution of the present paper is the following theorem,
which shows  that for ULC random variables a one-sided equivalent of 
(\ref{eq:vscaling})
holds. This result is proved in Appendix
\ref{sec:isop}, using a semigroup designed to preserve entropy, a development of
techniques in \cite{johnson21}.
We refer to this result as
the Restricted  Thinned Entropy Power Inequality (RTEPI), since it is a 
special case of the Thinned Entropy Power Inequality (\ref{eq:tepi}).
\begin{theorem}[RTEPI] \label{thm:rtepi}
Given any ULC random variable $X$, 
$$
 V(T_\alpha X) \geq \alpha V(X), \mbox{\;\;\; for  any $0 \leq \alpha \leq 1$.}$$
\end{theorem}

In the continuous case, the equivalents of Theorems \ref{thm:teci}
and \ref{thm:rtepi} allowed the full EPI, Theorem \ref{thm:epi}, 
to be deduced.
Despite this, in Section \ref{sec:failures} we describe how
 two apparently natural equivalents 
of the EPI, namely  (\ref{eq:firstepi}) and (\ref{eq:tepi}),
 in fact fail in general. These results are stated as Example
\ref{ex:fail1} and \ref{ex:fail2} respectively. In Theorem 
\ref{thm:tepis} we discuss some conditions
under which these results do hold.

However we can prove a discrete analogue of the rephrased
Entropy Power Inequality, Corollary \ref{cor:epilike}. 
The key operation is to invert the thinning operation
$T_\alpha$ on $X$, to create random variables $X^*$. 
This additional restriction
means that the result holds in less generality 
than Corollary \ref{cor:epilike}.
\begin{theorem} \label{thm:epilike}
 Given independent ULC random variables $X$ and $Y$, suppose
there exist $X^*$ and $Y^*$ such that $X = T_{\alpha} X^*$ and $Y = T_{1-\alpha}
Y^*$ for some $0 < \alpha < 1$, and such that $H(X^*) = H(Y^*)$. Then  
\begin{equation} \label{eq:epirephrase2} H(X+Y) \geq H(X^*),\end{equation}
with equality if and only if $X$ and $Y$ are Poisson.
\end{theorem}
\begin{IEEEproof} 
In analogy with the proof of Corollary \ref{cor:epilike},
for any $\alpha$ we  define $X_{\alpha}^*$ and $Y_{\alpha}^*$ (if 
such random variables exist) such that $X = T_\alpha X_\alpha^*$
and $Y = T_{1-\alpha} Y_{\alpha}^*$.
The Thinned Entropy Concavity Inequality, Theorem \ref{thm:teci}, implies
that
\begin{eqnarray}
H(X+Y) & = & H(T_{\alpha} X_{\alpha}^* + T_{1-\alpha} Y_{\alpha}^*) \nonumber \\
& \geq & \alpha H(X_\alpha^*) + (1-\alpha) H(Y_{\alpha}^*). \label{eq:todo}
\end{eqnarray}
This bound will hold for any $\alpha$, so
choosing $\alpha$ such that $H(X_{\alpha}^*) = H(Y_{\alpha}^*)$, we 
deduce the result.
\end{IEEEproof}

Unlike the continuous case, in general we cannot prove that this
is the right choice of $\alpha$, by optimizing 
  (\ref{eq:todo}).
However, we can give a related bound which we optimize, giving an
alternative heuristic
as to the right value of $\alpha$ to choose. 
That is, by Theorem
\ref{thm:rtepi} we deduce that
\begin{eqnarray} 
\lefteqn{ \alpha H(X_\alpha^*) + (1-\alpha) H(Y_{\alpha}^*) } \nonumber \\
& \leq & \alpha \EE \left( \frac{ V(X)}{\alpha} \right) + 
(1-\alpha) \EE \left( \frac{ V(Y)}{1-\alpha} \right).
\label{eq:altern}
\end{eqnarray}
Because $\EE(\cdot)$ is concave, the RHS of (\ref{eq:altern}) is maximized by
$\alpha = V(X)/(V(X) + V(Y))$. 

Note that it is not
always possible to find $X^*$ and $Y^*$ as 
required in Theorem \ref{thm:epilike}. For example, 
for $X \sim \bern(p)$,
there only exists $X^*$ such that $X = T_\alpha X^*$ when $\alpha \geq p$.
In general, for any random variable $X$ with support on $\{ 0,
\ldots, L \}$, there does
not exist $X^*$ such that $X = T_\alpha X^*$ for $\alpha < \ep X/L$
(since thinning preserves the support, then
$L \geq \ep X^* = \ep X/\alpha$).
Such an $X^*$ will exist for all $\alpha$ when  $X$ lies 
in certain parametric families,
including the geometric and Poisson, since these are
preserved by thinning (see \cite{johnson26}).

Some examples illustrate the bounds of Theorem
\ref{thm:epilike}:
\begin{example} Using Theorem \ref{thm:epilike}:

\begin{enumerate}
\item Given $X \sim \Pi_{\lambda}$ and $Y \sim \Pi_{\mu}$,  take
$X^* \sim Y^* \sim \Pi_{\mu+\lambda}$ and $\alpha = \lambda/(\lambda + \mu)$,
to confirm that equality does indeed hold in  (\ref{eq:epirephrase2})
in this case.

\item Given $X \sim \bino(n,p)$ and $Y \sim \bino(n,q)$, if $p + q \leq 1$
then 
 choosing $X^*
\sim Y^* \sim \bino(n,p+q)$ and $\alpha = p/(p+q)$, we deduce that
\begin{equation} \label{eq:binoineq}
 H( \bino(n,p) + \bino(n,q) ) \geq H( \bino(n, p+q)).\end{equation}
  By results in Poisson approximation, 
we expect that this inequality will be tightest for $n$ large and $p,q$
small. This result (\ref{eq:binoineq}) also follows from Theorem
1 of Shepp and Olkin \cite{shepp}, which states that if vector $\vc{p}$ majorizes
$\vc{q}$ then $H(B_{\vc{p}}) \leq H(B_{\vc{q}})$, where $B_{\vc{p}}$ is
the Bernoulli sum $\sum_{i=1}^n \bern(p_i)$.
Vector $(p+q, p+q, \ldots, p+q, 0,0, \ldots 0)$ majorizes vector
$(p,p, \ldots, p, q,q, \ldots q)$.

\item Given any identically distributed ULC random variables $X$ and $Y$, 
choosing $\alpha = 1/2$, we deduce that if there exists $X^*$ such that
$X = T_{1/2} X^*$ then
$$ H(X+Y) \geq H(X^*).$$
Note that such an $X^*$ does not exist for the random variables in 
Example \ref{ex:fail1}, which may be relevant to the fact that these
provide a counterexample to  (\ref{eq:firstepi}).
\end{enumerate}
\end{example}
\section{Monotonicity results} \label{sec:monotonicity}

The other major contribution
of this paper is to establish a monotonicity result in Theorem
\ref{thm:hmon}, which we regard as a 
discrete analogue of Artstein et al.'s Theorem
\ref{thm:hmonotone}.

In \cite{yu2}, corresponding monotonicity
results were proved regarding the
entropy and relative entropy of sums of thinned random variables, a situation
in which the two types of monotonicity are not equivalent.
Write $D(X) = D( X \| \Pi_{\lambda_X})$ for the relative entropy between
a random variable $X$ with mean $\lambda_X$ 
and a Poisson with the same mean.
Theorems 2 and 3 respectively of \cite{yu2} showed that  for 
independent and identically distributed $X_i$:
\begin{enumerate}
\item
the relative entropy
$D \left( \sum_{i=1}^{n} T_{1/n} X_i \right)$ is monotone decreasing in $n$,
\item for ULC $X_i$ the entropy
$H \left( \sum_{i=1}^{n} T_{1/n} X_i \right)$ is monotone increasing in $n$.
\end{enumerate}
In the spirit of Theorem \ref{thm:hmonotone}, we will place these
results from \cite{yu2} in a context where they can be deduced
from more general results, Lemma \ref{lem:dsub} and Theorem \ref{thm:hmon}. 
As a consequence we
give a proof of monotonicity of entropy
which uses distinct ideas from the convex ordering techniques used in \cite{yu2}.
The monotonicity of relative entropy is in fact implied by a stronger result
which is implicit in \cite{yu2}.
\begin{lemma} \label{lem:dsub}
Given positive $\alpha_i$ such that $\sum_{i=1}^{n+1} \alpha_i = 1$,
and writing $\alpha^{(l)} = \sum_{i \neq l} \alpha_i = 1 -\alpha_l$, then
for any independent $X_i$,
$$ n D \left( \sum_{i=1}^{n+1} T_{\alpha_i} X_i \right)
\leq \sum_{l=1}^{n+1} \alpha^{(l)} D \left( \sum_{i \neq l} T_{\alpha_i/\alpha^{(l)}} 
X_i \right).$$
\end{lemma}
\begin{IEEEproof}
Theorem 5 of \cite{yu2} shows that for independent random variables $Y_i$,
$$ n D \left( \sum_{i=1}^{n+1} Y_i \right) \leq \sum_{j=1}^{n+1} D
\left( \sum_{i \neq j} Y_i \right),$$
and Lemma 1 of \cite{yu2} shows that $D( T_\alpha X) \leq \alpha D(X)$.
Combining these two results we deduce that
\begin{eqnarray*}
n D \left( \sum_{i=1}^{n+1} T_{\alpha_i} X_i \right) 
& \leq &
\sum_{l=1}^{n+1} D \left( \sum_{i \neq l} T_{\alpha_i} X_i \right)  \\
& = & \sum_{l=1}^{n+1} D \left( T_{\alpha^{(l)}} \left( \sum_{i \neq l} 
T_{\alpha_i/\alpha^{(l)}} 
X_i \right) \right) \\
& \leq &
\sum_{l=1}^{n+1} \alpha^{(l)} D \left( \sum_{i \neq l} T_{\alpha_i/\alpha^{(l)}} 
X_i \right),
\end{eqnarray*}
and the result follows.
\end{IEEEproof}

We have to work harder to show that Theorem
\ref{thm:hmon}, the corresponding result in terms of
entropy, holds as well. The proof of this result is given in Appendix
\ref{sec:proofmon}.
\begin{theorem} \label{thm:hmon}
Given positive $\alpha_i$ such that $\sum_{i=1}^{n+1} \alpha_i = 1$,
and writing $\alpha^{(l)} = \sum_{i \neq l} \alpha_i$, then
for any independent ULC $X_i$,
\begin{equation} \label{eq:hmon} n H \left( \sum_{i=1}^{n+1} T_{\alpha_i} X_i \right)
\geq \sum_{l=1}^{n+1} \alpha^{(l)} H\left( \sum_{i \neq l} T_{\frac{\alpha_i}{\alpha^{(l)}}} 
X_i \right).\end{equation}
\end{theorem}
This result gives further support to the `general conjecture' of Gnedenko and Korolev
\cite[Pages 211--2]{gnedenko2} that `the universal principle of non-decrease of uncertainty
manifests itself in probability in the form of limit theorems when the limit is
taken with respect to infinitely increasing number of ``atomic'' random variables
involved in a model'. In particular Gnedenko and Korolev \cite[Page 215]{gnedenko2}
suggest that it is an important problem to `give information proofs of
limit theorems \ldots on convergence of random sums'. We believe that the fact that
thinning is an operation defined via random summation means that 
Theorem \ref{thm:hmon} represents progress in the direction proposed by these authors.

Note that Theorem \ref{thm:hmon} is a strengthened form of  Theorem
\ref{thm:teci},
indeed  Theorem \ref{thm:teci} can be deduced from it by successive deletion of
terms. 

Just as Theorem \ref{thm:teci} led to a 
proof of the rephrased Entropy Power Inequality
Theorem \ref{thm:epilike}, Theorem \ref{thm:hmon} leads to a strengthened version of Theorem
\ref{thm:epilike}, analogous to (\ref{eq:rephraseepi})
\begin{theorem} \label{thm:discepilike}
Assume there exist $Y_i^*$ and $\alpha_i$ such that $Y_i
= T_{\alpha_i} Y_i^*$ for each $i$, and there exists
some constant $H^*$ so
that 
the entropies satisfy
$H( \sum_{i \neq j} T_{\alpha_i/\alpha^{(j)}} Y_i^*) = H^*$ for
all $j$. Then
$$ H \left( \sum_{i=1}^{n+1} Y_i \right) \geq H^*.$$
\end{theorem}
\begin{IEEEproof}
Theorem \ref{thm:hmon} implies that
\begin{eqnarray*} n H \left( \sum_{i=1}^{n+1} Y_i \right) & = &  n H
\left( \sum_{i=1}^{n+1} T_{\alpha_i} Y_i^* \right) \\
& \geq & \sum_{l=1}^{n+1} \alpha^{(l)} H\left( \sum_{i \neq l} T_{\alpha_i/\alpha^{(l)}}
Y_i^* \right) \\
& = & n H^*,
\end{eqnarray*}
giving a discrete version
of the rephrased strengthened Entropy Power Inequality, 
(\ref{eq:rephraseepi}). \end{IEEEproof}

\section{Two natural discrete EPIs fail} \label{sec:failures}

Since the Poisson distribution shares with the Gaussian the property of infinite divisibility, as in \cite{kontoyiannis3} one natural analogue of Theorem \ref{thm:epi} 
comes from replacing $v$ by $V$, 
with equality holding if and only if $X$ and $Y$ are
Poisson. However,
as a counterexample provided by an anonymous referee previously
showed, such a result turns out not to be true. 
\begin{example} \label{ex:fail1}
For independent discrete random variables $X$ and $Y$,
it is not always the case that  
\begin{equation} \label{eq:firstepi} V(X+Y) \geq V(X) + V(Y). \end{equation}
A counterexample is that $X \sim Y$, 
$P_X(0) = 1/6$, $P_X(1) = 2/3$, $P_X(2) = 1/6.$ 
Notice that these $X$ and $Y$ are the sum of Bernoulli random variables,
and thus restriction of $X$ and $Y$ to the ULC class does not help.
\end{example}

 (\ref{eq:vscaling}) shows that an equivalent form of the EPI Theorem
\ref{thm:epi} is that for any $0 \leq \alpha \leq 1$,
\begin{equation} \label{eq:entpowercon}
v(\sqrt{\alpha} X + \sqrt{1-\alpha} Y) \geq \alpha v(X) + (1-\alpha) 
v(Y).
\end{equation}
(see \cite{dembo}). In analogy with this,
 we might make another conjecture, which again
turns out to not hold. 
\begin{example} \label{ex:fail2}
A natural conjecture, which we refer to as the 
Thinned Entropy Power Inequality,
 is that for independent discrete ULC random variables $X$ and $Y$,
for any $0 \leq \alpha \leq 1$,
\begin{equation} \label{eq:tepi}
V(T_\alpha X + T_{1-\alpha} Y) \geq \alpha V(X) + (1-\alpha)
V(Y),  \end{equation}
with equality for $0 < \alpha < 1$ if and only if $X$ and $Y$ are Poisson. 

However, taking $X \sim \bern(1/3) + \Pi_1$ and $Y \sim \Pi_{1000}$ and
$\alpha = 0.999$,  (\ref{eq:tepi}) is false. 
That is (taking all logs to base 2)
 $H(X) = 2.08286 \ldots$, and $V(X) = 1.27189\ldots$.
Clearly $V(Y) = 1000$.
Hence the RHS $= \alpha V(X) + (1-\alpha) V(Y) = 2.27062\ldots.$
Then $T_\alpha X + T_{1-\alpha} Y \sim \bern(\alpha/3) + \Pi_{\alpha + (1-\alpha) 1000}$, with $H(T_\alpha X + T_{1-\alpha} Y) = 2.55729\ldots$,
and $V(T_\alpha X + T_{1-\alpha} Y) = 2.25374\ldots$.
In this case  (\ref{eq:tepi})  fails.
\end{example}

Notice that  (\ref{eq:tepi}) fails
 even in the restricted case where $Y$ is Poisson, a case where we might
hope that even stronger results might hold, in analogy with work of Costa \cite{costa}.
The same is true of the conjecture (\ref{eq:firstepi}) -- if that result held for
$Y$ Poisson, then using Theorem \ref{thm:rtepi} would imply that 
(\ref{eq:tepi}) held in the same case.

As previously described, in the continuous case \cite{dembo}
 proves   (\ref{eq:entcon}) is equivalent to the
Entropy Power Inequality. The key fact in this proof is the scaling 
result, 
(\ref{eq:vscaling}). Since
Theorem \ref{thm:rtepi} is a one-sided version of this fact, we 
combine it with Theorem \ref{thm:teci} to
obtain the following partial results,
which were proved as Proposition 2 and Corollary 2 respectively of
 \cite{johnsonc6}, conditionally on the then unproved Theorem \ref{thm:rtepi},
so now hold without qualification.
\begin{theorem} \label{thm:tepis}
Consider independent ULC random variables $X$ and $Y$. 
  \begin{enumerate} 
\item
For any $\beta$, $\gamma$ such that 
$ \frac{\beta}{1-\gamma} \leq \frac{V(Y)}{V(X)} \leq
\frac{1-\beta}{\gamma}$ (note that in this case $\beta + \gamma <1$ unless
$V(X) = V(Y)$),
then
$$ V(T_\beta X + T_{\gamma} Y) \geq \beta V(X) + \gamma
V(Y).$$
\item If $Y \sim \Pi_{\mu}$, with $\mu \leq V(X)$, then
for all $0 \leq \alpha \leq 1$, 
$$V(T_\alpha X + T_{1-\alpha} Y) \geq \alpha V(X) + (1-\alpha)
V(Y).$$
\end{enumerate}
\end{theorem} 

We conjecture that there exist some $\alpha_- = \alpha_-(X,Y)$ and
$\alpha_+ = \alpha_+(X,Y)$ (perhaps defined 
in terms of the means and entropies of $X$ and $Y$)
such that for
$\alpha_- \leq \alpha \leq \alpha_+$,  (\ref{eq:tepi}) holds. However,
as Example \ref{ex:fail2} shows, the unrestricted version of this equation
fails.

It is worth noticing that the condition on $\beta$ and $\gamma$ in Theorem
\ref{thm:tepis}.1) can be restated as $\beta V(X) + (1-\gamma) V(Y)
\leq \min( V(X), V(Y))$. Hence by assuming a weaker bound,
 this theorem proves a stronger one.

\appendices

\section{Proof of RTEPI Theorem \ref{thm:rtepi}} 
\label{sec:isop}
We prove the Restricted Thinned Entropy Power Inequality, 
Theorem \ref{thm:rtepi}, using a
quantity $L(X)$ that plays a role analogous to the Fisher information in the work
of Blachman \cite{blachman} and Stam \cite{stam}.
\begin{definition}
For a random variable $X$ with
 probability mass function $P$, define the quantity
$$ L(X) = \sum_{z=0}^{\infty} (z+1) P(z+1) \log \left( 
\frac{P(z)}{P(z+1)} \right).$$
\end{definition}
We develop the argument in \cite{johnson21}, where we adapted random variables
by thinning and then adding an independent Poisson random variable:
\begin{definition} \label{def:umap} \mbox{ } For a positive function $f(\alpha)$,
 define the combined map $U_{\alpha,f(\alpha)}$ that thins and then
adds an independent Poisson random variable:
$$ U_{\alpha,f(\alpha)} X = T_\alpha X + \Pi_{f(\alpha)}.$$
\end{definition}
For most of this section, we assume that the random variable $X$ has finite
support.
\begin{proposition} \label{prop:heateqn}
Consider a continuously differentiable function $f$ with
$f(1) = 0$. Assume either
(a) $f(t) \equiv 0$ for all $t$ or (b)
$f(t) > 0$ for $t < 1$.
Given ULC $X$ with finite support, 
writing $X_t = U_{t,f(t)} X$ 
and $P_{t}(z) = \pr(X_t= z)$, then for any $0 < t < 1$
$$ \frac{\partial}{\partial t} H(X_t) = \frac{
L(X_t)}{t}
- r(t) \sum_{z=0}^{\infty} P_{t}(z) \log  
\frac{P_{t}(z)}{P_t(z+1)},$$
where $r(t) = f(t)/t - f'(t)$. Equivalently, 
$f(t) = t f(1) + t \int_{t}^1 r(\beta)/\beta d \beta$.
\end{proposition}
\begin{IEEEproof}
From Equation (8) of \cite{johnson21}, we know that the mass function of $X_t$
satisfies
\begin{equation} \label{eq:heateqn2}
 \frac{\partial }{\partial t} P_{t}(z)
= \Delta^* \left( \frac{(z+1) P_{t}(z+1)}{t}
- r(t) P_{t}(z) \right),\end{equation}
where adjoint operators $\Delta$ and $\Delta^*$ are defined by 
$\Delta^* g(x) = g(x-1) - g(x)$ and $\Delta g(x) = g(x+1) - g(x)$.
Then we simply differentiate the entropy, using  (\ref{eq:heateqn2}) to obtain
\begin{eqnarray*}
\lefteqn{ \frac{\partial}{\partial t} H(P_t) } \\
& = & - \sum_{z=0}^{\infty} \frac{ \partial P_{t}}{\partial t}(z)
\log P_{t}(z) - \sum_{z=0}^{\infty} \frac{ \partial P_{t}}{\partial t}(z) \\
& = & - \sum_{z=0}^{\infty} \Delta^* \left( \frac{(z+1) P_{t}(z+1)}{t}
- r(t) P_{t}(z) \right) \log P_{t}(z) \\
& = & \sum_{z=0}^{\infty} \left( \frac{(z+1) P_{t}(z+1)}{t}
- r(t) P_{t}(z) \right) \log \frac{P_{t}(z)}{P_{t}(z+1)} 
\end{eqnarray*}
and the result follows, where this final step uses Fubini's theorem.

The differentiation of the infinite series at $t$ can be justified 
in the case (a) since then the sum is simply a finite one. In case (b)
it can be justified by a result
(see \cite{porter}) concerning $H(s) = \sum_{z=0}^{\infty}
u_s(z)$ with $a \leq s \leq b$. The 
derivative $\frac{\partial H}{\partial s}
= \sum_{z=0}^{\infty} \frac{\partial }{\partial s}
u_s(z)$ for $a < s < b$, assuming
that $\frac{\partial }{\partial s} u_s(z)
$ exist, and are uniformly bounded as $\left| \frac{\partial }{\partial s}
u_s(z) \right|
 \leq M(z)$, for
all $a < s < b$, where $\sum_{z=0}^{\infty} M(z) < \infty$. 

Given a particular $0 < t < 1$, we can choose $a < t < b$ such that this 
result holds.
In this case, writing $\lambda  = \ep X$, Equation (9) of \cite{johnson26}
shows that $\pr(T_s X = 0) \geq (1-s)^{\lambda}$, so that
\begin{equation} \label{eq:pszlbd}
 P_s(z) \geq \pr(T_s X = 0) \pr( \Pi_{f(s)} = z) \geq
(1-s)^{\lambda} \frac{e^{-f(s)} f(s)^z}{z!},\end{equation}
hence for $a < s < b$, for all $z$, we can bound
$$ | -\log P_s(z) | \leq - \lambda \log(1-s) + f(s) + z |\log f(s)| 
+ \log z!.$$
Sincce $f(s)$ is continuous and bounded away from zero on $(a,b)$,
Stirling's formula means that this can be uniformly
bounded by $C_1 + C_2 z^2$, where $C_1$ and $C_2$ depend on $a$ and $b$.

Similarly, the triangle inequality means that
\begin{eqnarray*} \left| \frac{\partial P_s}{\partial s}(z) \right|
& \leq &  \frac{z P_s(z)}{s} + |r(s)| P_s(z-1) \\
& &  + \frac{(z+1) P_s(z+1)}{s}
+ |r(s)| P_s(z),
\end{eqnarray*}
so the fact that $X$, and hence $X_s$, is ULC means that $P_s(z)
\leq (P_s(1)/P_s(0))^z/z! P_s(0)$. Hence, since (\ref{eq:pszlbd})
means that the ratio $P_s(1)/P_s(0)$ is uniformly bounded on $(a,b)$,
the result follows by continuity (and hence boundedness) of $r(t)$.
 
Note that although this result is stated for ULC $X$
with finite support, it should hold for any
random variables such that the differentiation step can be justified.
\end{IEEEproof}
Writing $\JJ(t) = \EE'(t) = \sum_{z=0}^{\infty} \Pi_t(z) \log ( (z+1)/t)$ (a positive function), we state the following isoperimetric inequality, equivalent to the RTEPI Theorem \ref{thm:rtepi}, a technique suggested by \cite{kontoyiannis3}.
This result may be of independent interest.
\begin{theorem} \label{thm:isop}
For all ULC random variables $X$ with finite support, 
$$ L(X) \leq V(X) \JJ( V(X)).$$
\end{theorem}
\begin{lemma} \label{lem:equiv} For random variables $X$ with finite 
support, Theorems \ref{thm:rtepi} and \ref{thm:isop} 
are equivalent. \end{lemma}
\begin{IEEEproof} 
Write $g(\alpha)$ for $V(T_\alpha X)$.
Assume Theorem \ref{thm:rtepi} holds, so that $g(\alpha) \geq \alpha g(1)$
or, rearranging, that for $\alpha <1$
$$ \frac{ g(\alpha) - g(1)}{\alpha - 1} \leq g(1),$$
(the change of direction of the inequality comes since $\alpha < 1$). Letting $\alpha
\rightarrow 1$, we see that the RTEPI implies that $g'(1) \leq g(1)$.

The key is to observe that using Proposition \ref{prop:heateqn}, the derivative 
of $H(T_\alpha X)$ with respect to $\alpha$ is $L(T_\alpha X)/\alpha$.
This means that by the chain rule the derivative 
\begin{eqnarray} 
g'(\alpha) & = & \left( \EE^{-1} \right)'( H(T_\alpha X)) \frac{L(T_\alpha X)}
{\alpha} \nonumber \\
& =  & \frac{L(T_\alpha X)}{\alpha \JJ( \EE^{-1}(H(T_\alpha X)))} \nonumber
\\
& =  & \frac{L(T_\alpha X)}{\alpha \JJ( V(T_\alpha X))}, \label{eq:gder}
\end{eqnarray}
so taking $\alpha =1$, the result $g'(1) \leq g(1)$ becomes Theorem \ref{thm:isop}.

We deduce the reverse implication by
using  (\ref{eq:gder}), and applying
Theorem \ref{thm:isop} to the random variable $T_\alpha X$, to deduce that
$$ g'(\alpha) = \frac{L(T_\alpha X)}{\alpha \JJ( V(T_\alpha X))}
\leq \frac{V(T_\alpha X)}{\alpha} = \frac{g(\alpha)}{\alpha}.$$
This implies that $g(\alpha)/\alpha$ is 
decreasing in $\alpha$, which means that $g(\alpha)/\alpha \geq g(1)/1$, which
is Theorem \ref{thm:rtepi}. 
\end{IEEEproof}
We prove Theorem \ref{thm:isop} next, and hence deduce that
Theorem \ref{thm:rtepi} holds by Lemma \ref{lem:equiv}.
Our approach involves the map $U_{\alpha,f(\alpha)}$ which preserves
the entropy (as opposed to preserving the mean as in \cite{johnson21}).

\begin{IEEEproof}[Proof of Theorem \ref{thm:isop}]
Since $L(X) = \frac{\partial H}{\partial \alpha}(T_\alpha X) |_{\alpha=1}$,
we know that $L(X)$ need not always be positive (consider for example
$X \sim \bern(p)$  with $p > 1/2)$. 
However,
note that if $L(X) \leq 0$, then automatically $L(X) \leq 0 \leq V(X) \JJ(V(X))$, 
as required. Hence, we can restrict our interest to the case where 
$L(X) > 0$.

Now, 
$H(T_\alpha X)$ is a positive concave function of $\alpha$ which (since 
by \cite{johnson21} it
is upper bounded by the entropy of a $\Pi_{\alpha \lambda_X}$ random variable) 
tends to
zero as $\alpha$ tends to zero. Hence, $H(T_\alpha X)$ can only be
decreasing in $\alpha$ for $\alpha \in (\alpha^*,1]$, for some 
$\alpha^* > 0$.
Hence, if $L(X) > 0$, then $L(T_\alpha X) \geq 0$ for all $\alpha
\in [0,1]$ and $H(T_\alpha X)$ is a increasing function of $\alpha$
for all $\alpha \in [0,1]$. Hence, it is possible to perform an
interpolation argument -- that is, we can find $f(t) \geq 0$ such 
that $X_t = U_{t,f(t)} X$ has constant entropy. We write $\lambda_t$ for the mean
of $X_t$.

This means
that, since the semigroup interpolates between $X_1 \sim X$ and $X_0 \sim \Pi_{\lambda'}$,
a Poisson random variable with mean $\lambda'$, we can deduce that
$$H(X) = H(X_1) = H(X_0) = H(\Pi_{\lambda'}) = \EE(\lambda'),$$ or that
$\lambda' = V(X)$.

Motivated by Proposition \ref{prop:heateqn}
we consider properties of
$r(t) = L(X_t)/\left(t \sum_{z=0}^{\infty} P_{t}(z) \log 
\left( \frac{P_{t}(z)}{P_t(z+1)} 
\right) \right)$. Note that by Chebyshev's rearrangement lemma (see for 
example Equation (1.7)
of \cite{kingman})
$$L(X_t) = \sum_{z=0}^{\infty} P_t(z) \left( \frac{(z+1) P_t(z+1)}{P_t(z)} \right) \log \left( \frac{P_{t}(z)}{P_t(z+1)} \right)$$ is the expectation of the product of an increasing and decreasing function,
so $L(X_t) \leq \lambda_t  \sum_{z=0}^{\infty} P_{t}(z) \log \left( \frac{P_{t}(z)}{P_t(z+1)} \right)$, or $r(t) \leq \lambda_t/t$.
We can write $L(X_t)$ as
\begin{eqnarray}
\lefteqn{ - \lambda_t D(P_t^{\#} \| P_t) +
\sum_{z=0}^{\infty} (z+1) P_t(z+1) \log \left( \frac{z+1}{\lambda_t} \right)} 
\nonumber \\
& \leq &  - D(P_t \| \Pi_{\lambda_t})  \nonumber \\
&  & +
\sum_{z=0}^{\infty} (z+1) P_t(z+1) \log \left( \frac{z+1}{\lambda_t} \right)
\label{eq:wu} \\
& = & H(X_t) - \sum_{z=0}^{\infty} P_t(z+1) \log (z+1)!  - \lambda_t 
\nonumber \\
& & + \sum_{z=0}^{\infty} (z+1) P_t(z+1) \log (z+1),
\label{eq:toderiv}
\end{eqnarray}
where $P^{\#}_t(x) = P_t(x+1) (x+1)/\lambda_t$ is the size-biased version of 
$P_t$,
and  (\ref{eq:wu}) follows by Equation (0.6) of Wu \cite{wu}.

Theorem \ref{thm:isop} will follow if we can prove that this expression (\ref{eq:toderiv}),
which we shall refer to as $U(X_t)$, is a decreasing
function of $t$.
That would mean that
\begin{eqnarray*}
L(X) & = & L(X_1) \leq U(X_1) \leq U(X_0) \\
& = & \lambda' \JJ(\lambda') = V(X) \JJ(V(X)).
\end{eqnarray*}
In fact, since $H(X_t)$ is constant, equivalently, we will prove that
$U(X_t) - H(X_t)$ is a decreasing function of $t$.

{\bf Case A:  $r(t) > 0$ for all $t$.}
We simply differentiate  (\ref{eq:toderiv}), using Equation (\ref{eq:heateqn2}),
and express $\frac{\partial U(X_t)}{\partial t}$ as  
\begin{eqnarray}
\lefteqn{
 \sum_{z=0}^{\infty} P_t(z+1) \left( \frac{ (z+2) P_{t}(z+2)}{t
P_t(z+1)} - r(t) \right) (z+1) \log \frac{z+2}{z+1} }
\nonumber \\ 
& & +
r(t) - \frac{\lambda_t}{t}. \hspace*{6cm} \;\;\;\; \label{eq:derivu} 
\end{eqnarray}
The term-by-term differentiation can be justified as before, since the
assumption that $r(t) = - (f(t)/t)' > 0$ implies that $f(t) > 0$ for $t < 1$,
so the
assumptions of Proposition \ref{prop:heateqn} hold. Hence the entropy can
indeed be differentiated, and the functions
$\log z!$ and $z \log z$ can be controlled using a similar argument.
Since
$-(z+1) \log \frac{z+2}{z+1} + 1 \geq 0$,  Equation (\ref{eq:derivu}) is
increased on replacing $r(t)$ by the (larger) value $\lambda_t/t$, so
we deduce that $\frac{\partial U(X_t)}{\partial t}$ is less than or equal to
\begin{equation}
 \sum_{z=0}^{\infty} \frac{P_t(z+1)}{t} \left( \frac{ (z+2) P_{t}(z+2)}{
P_t(z+1)} - \lambda_t \right) (z+1) \log \frac{z+2}{z+1}. \label{eq:derivv}
\end{equation}
Observe that (\ref{eq:derivv}) is the covariance of decreasing and increasing functions, and
hence is negative by the Chebyshev rearrangement lemma.
We have shown that if $L(X_t) > 0$ for all $t$, so that $r(t) > 0$ for all $t$, then
$L(X_t)$ is a decreasing function at $t$.

{\bf Case B: $r(t) \leq 0$ for some t.} 
Recall that we need only consider the case 
where $L(X) = L(X_1) > 0$. Define
$t^* = \sup \{ t \geq 0: r(t) \leq 0 \}$. 
Suppose that $t^* > 0$. For all $t > t^*$, $r(t) > 0$, so that 
for all $t > t^*$, we know that $L(X_t) \geq L(X) > 0$. By
considering $t$ arbitrarily close to $t^*$, continuity of 
$L(X_t)$ implies that $L(X_t) > 0$ for all $t \in (t^*-\epsilon,
t^*)$. This contradicts the assumption that $t^* > 0$, so we
deduce that $r(t) > 0$ for all $t > 0$, and the result follows. 
\end{IEEEproof}

\begin{IEEEproof}[Proof of Theorem \ref{thm:rtepi}]
By Lemma \ref{lem:equiv} we deduce from Theorem \ref{thm:isop} 
that the RTEPI, 
Theorem \ref{thm:rtepi} holds for ULC $X$ with finite support.

For general ULC $X$, let $X^{(k)}$ be the random variable $X$ truncated
at $k$, for $k=1, 2, \ldots$. Then the mass function of $T_\alpha X^{(k)}$ 
tends
to that of $T_\alpha X$ pointwise, for all 
$0< \alpha \leq 1$.  Moreover, the mean of $T_\alpha X^{(k)}$
tends to that of $T_\alpha X$.  The argument of Part 2) in Theorem 1 
of \cite{yu2}
shows that $H(T_\alpha X^{(k)}) 
\rightarrow H(T_\alpha X)$ as $k \rightarrow \infty$ (the basic
argument is to apply Fatou's lemma twice).  Because $\EE^{-1}(.)$ is
continuous, we have $V(T_\alpha X^{(k)}) 
\rightarrow V(T_\alpha X)$ as $k \rightarrow \infty$.  Thus
Theorem \ref{thm:rtepi} 
holds by taking a limit on the finite support result.
\end{IEEEproof}
\section{Proof of monotonicity Theorem \ref{thm:hmon}} \label{sec:proofmon}
In this section, we prove monotonicity of entropy by analysing certain directional
derivatives of an `energy' functional $\Lambda$.
For $X$ with expectation $\lambda_X$, we write
$\Lambda(X) = - \ep \log \Pi_{\lambda_X} (X)
= \lambda_X + \ep \log X! - \lambda_X \log \lambda_X$. In this section,
we will establish the following proposition:
\begin{proposition} \label{prop:ent}
Given positive $\alpha_i$ such that $\sum_{i=1}^{n+1} \alpha_i = 1$,
and writing $\alpha^{(l)} = \sum_{i \neq l} \alpha_i$, then
for any independent ULC $X_i$, 
\begin{equation} \label{eq:lammon}
 n \Lambda \left( \sum_{i=1}^{n+1} T_{\alpha_i} X_i \right)
\geq \sum_{l=1}^{n+1} \alpha^{(l)} \Lambda \left( \sum_{i \neq l} T_{\alpha_i/\alpha^{(l)}} 
X_i \right).\end{equation}
\end{proposition}

As in \cite{johnsonc6}, Lemma \ref{lem:dsub} can be subtracted from Proposition \ref{prop:ent} to deduce that Theorem \ref{thm:hmon} holds.
We will write $\vd{\alpha} = (\alpha_1, \ldots, \alpha_{n+1})$ and 
given independent ULC $X_i$ with means $\lambda_i$ we will define
the function
$\Phi(\vd{\alpha}) = \Lambda \left( \sum_{i=1}^{n+1} T_{\alpha_i} X_i \right)$.
We write $\pr_{\vd{\alpha}}(\vc{x}) = 
\pr \left( T_{\alpha_1} X_1 = x_1, \ldots,  T_{\alpha_{n+1}} X_{n+1} = 
x_{n+1} \right)$ and $\qr_{\vd{\alpha}}( s) = \sum_{ \vc{x}: \sum_i x_i = s} \pr_{\vd{\alpha}}(\vc{x})$.
In order to establish Proposition \ref{prop:ent}, we will need to understand
the properties of the Hessian matrix $\Phi''$, which we write as the 
sum of two matrices $\Phi'' = \Phi''_1 + \Phi''_2$. The first matrix,
$$ \Phi''_1(\vd{\alpha})_{ij} = \frac{\partial^2}{\partial \alpha_i \partial \alpha_j}  \sum_{s=0}^{\infty} \qr_{\vd{\alpha}}( s) \log s!,$$
can be evaluated using 
Equation (\ref{eq:heateqn2}) -- we omit the details for brevity:
\begin{lemma} \label{lem:hess1}
For any $\vd{\alpha}$, $i$ and $j$ the
derivative
\begin{equation}
\Phi''_1(\vd{\alpha})_{ij} =
\sum_{s=0}^{\infty}
\sum_{ \vc{x}: \sum_i x_i = s} \pr_{\vd{\alpha}} (\vc{x}) 
\frac{ x_i (x_j-\delta_{ij})}{\alpha_i^2}
\log \left( \frac{s}{s-1} \right) 
\end{equation}
\end{lemma}

The second term of the Hessian, $\Phi''_2$, can be explicitly evaluated
by writing $\theta(t) = t -t \log t$ and expressing
\begin{eqnarray}
 \Phi''_2(\vd{\alpha})_{ij} & = &
\frac{\partial^2}{\partial \alpha_i \partial \alpha_j}
\theta \left(\sum_{k=1}^{n+1} \alpha_k \lambda_k \right)  \nonumber \\
& = & -\frac{ \lambda_i \lambda_j}{\sum_{k=1}^{n+1} \alpha_k \lambda_k}. \label{eq:hess2}
\end{eqnarray}
We now examine the Hessian $\Phi''$ in more detail, using techniques that
extend the proof of Theorem \ref{thm:teci} given in \cite{johnsonc6}, first
introducing a sufficient condition. 
\begin{condition} \label{cond:suff}
We say that vectors $\vd{\mu}$ and $\vd{\beta}$ satisfy the positive splitting condition
if there exist positive $u_{ij}$ such that
\begin{enumerate}
\item For all $i,j$ the terms $$u_{ij} + u_{ji} = v_{ij}(\vd{\beta}, \vd{\mu})
:= \left( \frac{\mu_i}{\beta_i} - \frac{\mu_j}{\beta_j} \right)^2 
\beta_i \beta_j \lambda_i \lambda_j.$$
\item For all $j$ the terms $\left( \sum_{i \neq j} u_{ij} \right)/(\beta_j 
\lambda_j)$ take the same value, $S$ say.
\end{enumerate}
\end{condition}
Observe that if Condition \ref{cond:suff} holds, then multiplying the terms in
Part 2.
by $\beta_j
\lambda_j$ and summing over $j$ we deduce that  
\begin{eqnarray*} S & = & \frac{\sum_{i < j} v_{ij}(\vd{\beta}, \vd{\mu})}{
\sum_k \beta_k \lambda_k} \\
& = & \frac{\sum_{i < j} \left( \mu_i/\beta_i - \mu_j/\beta_j \right)^2 \beta_i \beta_j \lambda_i \lambda_j}{
\sum_k \beta_k \lambda_k} \\
& = & \frac{ - \sum_{i \neq j} \mu_i \mu_j \lambda_i \lambda_j + \sum_i
(\mu_i^2 \lambda_i/\beta_i) 
\left( \sum_{j \neq i} \beta_j \lambda_j \right)}{
\sum_k \beta_k \lambda_k},
\end{eqnarray*}
so that
\begin{eqnarray} 
\lefteqn{ \left( \sum_i
\frac{\mu_i^2 \lambda_i}{\beta_i} \right) - S } \nonumber \\
& = & \frac{1}{\sum_k \beta_k \lambda_k}
\left( \sum_i
\frac{\mu_i^2 \lambda_i}{\beta_i} (\beta_i \lambda_i) +
 \sum_{i \neq j} \mu_i \mu_j \lambda_i \lambda_j \right) \nonumber \\
& = & \frac{ \left( \sum_k \mu_k \lambda_k \right)^2}
{\sum_k \beta_k \lambda_k}.
\label{eq:lminus} \end{eqnarray}
This property allows us to deduce the following result:
\begin{theorem} \label{thm:suff}
If $\vd{\mu}$ and $\vd{\beta}$ satisfy the positive splitting condition,
Condition \ref{cond:suff},  then $\vd{\mu}^T
\Phi''(\vd{\beta}) \vd{\mu} \leq 0$.
\end{theorem}
\begin{IEEEproof}
We use Lemma \ref{lem:hess1} to deduce that,
writing $\vc{e}_i$ for the $i$th unit vector, $s = \sum_i x_i$ and 
$\vc{x}^{(i,-)} = \vc{x} - \vc{e}_i$, then we can express the 
product $\vd{\mu}^T \Phi''_1(\vd{\beta}) \vd{\mu}$ as
\begin{eqnarray}
\lefteqn{ 
\sum_{ \vc{x}} \pr_{\vd{\beta}} (\vc{x}) \sum_{i=1}^{n+1} \left[ 
\frac{ \mu_i^2 x_i (x_i-1)}{\beta_i^2}
+ \sum_{j \neq i}
 \frac{\mu_i \mu_j x_i x_j}{\beta_i \beta_j} \right]
\log \left( \frac{s}{s-1} \right) } \nonumber \\
&=  & \sum_{i=1}^{n+1} 
\sum_{ \vc{x} } \pr_{\vd{\beta}} (\vc{x}) x_i 
\log \left( \frac{s}{s-1} \right)  \nonumber \\
& & \hspace*{1cm} \times \left[ 
\frac{ \mu_i^2}{\beta_i^2} \left( \sum_k x_k - 1 \right)
- \sum_{j \neq i} \frac{u_{ij} x_j}{\beta_i \beta_j \lambda_i
\lambda_j} \right] \hspace*{1cm} \label{eq:suff1} \\
& \leq  & \sum_{i=1}^{n+1} 
\sum_{ \vc{x} } \beta_i \lambda_i \pr_{\beta} (\vc{x}^{(i,-)}) 
\log \left( \frac{s}{s-1} \right)
 \nonumber \\
& & \hspace*{1cm} \times \left[ 
\frac{ \mu_i^2}{\beta_i^2} \left( s - 1 \right)
- \sum_{j \neq i} \frac{u_{ij} x_j}{\beta_i \beta_j \lambda_i
\lambda_j} \right]
 \label{eq:suff2} \\
& = &  \sum_{s=0}^{\infty} \qr_{\vd{\beta}}(s) s \log \left( \frac{s+1}{s} \right)
 \left[ \left( \sum_{i=1}^{n+1} 
\frac{ \mu_i^2 \lambda_i}{\beta_i} \right) - S \right]
 \label{eq:suff3} \\
& \leq &  \frac{ \left( \sum_k \mu_k \lambda_k \right)^2}
{\sum_k \beta_k \lambda_k} = - \vd{\mu}^T \Phi''_2(\vd{\beta}) \vd{\mu}. \label{eq:suff5}
\end{eqnarray}
Here Equation (\ref{eq:suff1}) follows by comparing coefficients of 
$x_i x_j$, using Part 1. of Condition \ref{cond:suff}.
Equation (\ref{eq:suff2}) follows as in \cite{johnsonc6}, using 
Chebyshev's rearrangement lemma, and the fact that
$(x_i + w) \log( (x_i + w)/(x_i + w -1))$ is increasing in $x_i$
and $\log( (x_i
 + w)/(x_i + w -1)$ is decreasing in $x_i$ (coupled
with the assumption that $u_{ij} \geq 0$). Equation (\ref{eq:suff3})
uses Part 2. of Condition \ref{cond:suff}.
Equation (\ref{eq:suff5}) follows using (\ref{eq:lminus}) since,
as in \cite{johnsonc6}, 
$s \log ((s+1)/s) \leq 1$. Finally we use the expression for $\Phi_2''$ given
in Equation (\ref{eq:hess2}). 
\end{IEEEproof}

We can use this result to complete the proof of monotonicity of
entropy, Theorem \ref{thm:hmon}, by proving Proposition \ref{prop:ent}.

\begin{IEEEproof}[Proof of Proposition \ref{prop:ent}]
For each $l$, we can define a one-parameter map which interpolates between
the values of $\vd{\alpha}$. That is, for each $l$, define
$$ \vc{A}_l(t) = (1-t) \vd{\alpha}^{(l)} + t \vc{e}_l,$$
where $\vd{\alpha}^{(l)} = ( \alpha_1, \ldots, 
\alpha_{l-1}, 0, \alpha_{l+1}, \ldots, \alpha_{n})/\alpha^{(l)}$ is
the renormalized `leave one out' vector, and
 $\vc{e}_l$ is the $l$th unit vector. We write $\vd{\mu}_l = 
\vc{e}_l - \vd{\alpha}^{(l)}
  = \frac{\partial}{\partial t} \vc{A}_l(t)$.
Observe that $\vc{A}_l(0) = 
\vd{\alpha}^{(l)}$ and $\vc{A}_l( \alpha_l) = \vd{\alpha}$,
meaning
that by Taylor's theorem,  
for some $t_l^* \in [0,\alpha_l]$, if the relevant Hessian term is negative,
\begin{eqnarray}
\Phi( \vd{\alpha}^{(l)}) - \Phi( \vd{\alpha}) & = &  \alpha_l \vd{\mu}_l^T 
\Phi'(\vd{\alpha}) + \frac{\alpha_l^2}{2} \vd{\mu}_l^T
\Phi''(\vc{A}_l(t_l^*) ) \vd{\mu}_l \nonumber \\
& \leq & \alpha_l \vd{\mu}_l^T 
\Phi'(\vd{\alpha}). \label{eq:tocheck}
\end{eqnarray}
 If this is
true for each $l$, on multiplying by
$\alpha^{(l)}$ and summing over $l$ we deduce that
$\sum_{l=1}^{n+1} \alpha^{(l)} \Phi( \vd{\alpha}^{(l)}) 
\leq n \Phi( \vd{\alpha})$, and
the proof is complete. (This uses the property that $\sum_l \alpha^{(l)}
\alpha_l \vd{\mu}_l = \vc{0}$, which is a consequence of the fact that
$\sum_l \alpha^{(l)} \alpha_l \vd{\alpha}^{(l)}
= \sum_l \alpha_l ( \alpha_1, \ldots, 
\alpha_{l-1}, 0, \alpha_{l+1}, \ldots, \alpha_{n}) = (\alpha_1 \alpha^{(1)},
\ldots, \alpha_{n+1} \alpha^{(n+1)}) = 
\sum_l \alpha^{(l)} \alpha_l \vc{e}_{l}$, as required).

We complete the proof by checking the negativity of the
relevant Hessians by testing positive splitting, Condition \ref{cond:suff},
and applying Theorem \ref{thm:suff}. There are considerable simplifications
in this case, since the majority of the values of $v_{ij}(\vc{A}_l(t_l^*),
\vd{\mu}_l)$
vanish. 
That is, if $i,j \neq l$ then for any $t$ the $v_{ij}(\vc{A}_l(t), \vd{\mu}_l)$
becomes
$$  \left( \frac{\alpha_i/\alpha^{(l)}}{\alpha_i 
(1-t)/
\alpha^{(l)} }  - \frac{\alpha_j/\alpha^{(l)}}{\alpha_j (1-t)/\alpha^{(l)} } \right)^2 
\alpha_i(t) \alpha_j(t) \lambda_i \lambda_j = 0.$$
In the remaining case, when $i \neq l$ and $j =l$, the
$v_{il}(\vc{A}_l(t), \vd{\mu}_l)$ is
\begin{equation}
 \left( \frac{\alpha_i/\alpha^{(l)}}{\alpha_i (1-t)/
\alpha^{(l)} }  + \frac{1}{t} \right)^2 
\frac{\alpha_i (1-t)}{\alpha^{(l)}}  t \lambda_i \lambda_l 
= \frac{ \alpha_i \lambda_i \lambda_l}{\alpha^{(l)} t (1-t)}.
\end{equation}
We can exhibit a set of positive solutions to the required
equations by writing $\lambda(t)
= \sum_i \alpha_i(t) \lambda_i$, $\lambda^{(l)}(t) =
\sum_{i \neq l} \alpha_i(t) \lambda_i = \lambda(t) - t \lambda_l$
 and $S = (\lambda^{(l)}(t) \lambda_l)/
(t(1-t)^2 \lambda(t))$.
Then define $u_{ij}$ to be zero unless $i$ or $j$ equals
$l$, in which case for $i \neq l$, 
\begin{equation} u_{li} = \frac{S \alpha_i \lambda_i (1-t)}{\alpha^{(l)}} \mbox{ and }
 u_{il}  =  \frac{ \lambda_l^2 \alpha_i \lambda_i}{(1-t) \alpha^{(l)} \lambda(t)}.
\label{eq:udefn}
\end{equation} 
We confirm that this choice of $u$ satisfies Condition \ref{cond:suff} --
firstly clearly these terms are positive. Secondly for all $i \neq l$, the sum
\begin{eqnarray*}
u_{li} + u_{il} & = & \frac{\alpha_i \lambda_i}{\alpha^{(l)}} \left( S (1-t) +
\frac{\lambda_l^2}{\lambda(t) (1-t)} \right) \\
& = & \frac{\alpha_i \lambda_i}{\alpha^{(l)}}
\left( \frac{\lambda_l}{t(1-t)} \right) = v_{il}( \vc{A}_l(t), \vd{\mu}_l).
\end{eqnarray*}
Finally, for $u$ as defined in  (\ref{eq:udefn}), writing
$A_{l,j}(t)$ for the $j$th component of $\vc{A}_l(t)$,
the sums $\sum_{i \neq j} u_{ij}/(A_{l,j}(t) \lambda_j)$ do 
indeed equal $S$ for
each $j$. Specifically, for $j \neq l$ there is only non-zero term in the sum,
giving $u_{lj}/(A_{l,j}(t) \lambda_j) = S$, since $A_{l,j}(t) = \alpha_j (1-t)/
\alpha^{(l)}$.
 For $j = l$, since $A_{l,j}(t) = t$,
the sum becomes
$$ \frac{\sum_{i \neq l} u_{il}}{A_{l,j}(t) \lambda_l} = 
\frac{ \lambda_l \left( \sum_{i \neq l} \alpha_i \lambda_i \right)}
{ t (1-t) \alpha^{(l)} \lambda(t)} = S,$$ 
as required. Hence Condition \ref{cond:suff} holds in this case, so we can apply Theorem \ref{thm:suff} to deduce that $\vd{\mu}_l^T \Phi''(\vc{A}_l(t)) \vd{\mu}_l
\leq 0$ for all $t$. This means that  
 (\ref{eq:tocheck}) holds for each $l$,
and the proof of Proposition \ref{prop:ent} is complete.
\end{IEEEproof}
\section*{Acknowledgements}
 The authors would like to thank Ioannis
Kontoyiannis and Peter Harremo\"{e}s for discussions concerning the discrete
Entropy Power Inequality, and in particular for some of
the notation used in this paper.



\begin{IEEEbiographynophoto}
{Oliver Johnson} received the BA degree in 1995, Part III Mathematics
in 1996 and a PhD in 2000, all from the University of Cambridge.
He was Clayton Research Fellow of Christ's College
and Max Newman Research Fellow of Cambridge University until 2006,
during which time he published the book {\em Information Theory and
the Central Limit Theorem} in 2004. Since 2006 he has been Lecturer
in Statistics at Bristol University.
\end{IEEEbiographynophoto}
\begin{IEEEbiographynophoto}
{Yaming Yu} (M'08) received the B.S. degree in mathematics from Beijing
University, Beijing, China, in 1999, and the Ph.D. degree in statistics
from Harvard University, Cambridge, MA, in 2005.  Since 2005 he has been
an Assistant Professor in the Department of Statistics at the University
of California, Irvine.
\end{IEEEbiographynophoto}

\end{document}